# Dipole-Dipole Interactions of Charged-Magnetic Grains

Jonathan Perry, Lorin S. Matthews, and Truell W. Hyde, *Member, IEEE*

*Abstract*— The interaction between dust grains is an important process in fields as diverse as planetesimal formation or the plasma processing of silicon wafers into computer chips. This interaction depends in large part on the material properties of the grains, for example whether the grains are conducting, non-conducting, ferrous or non-ferrous. This work considers the effects that electrostatic and magnetic forces, alone or in combination, can have on the coagulation of dust in various environments. A numerical model is used to simulate the coagulation of charged, charged-magnetic and magnetic dust aggregates formed from ferrous material and the results are compared to each other as well as to those from uncharged, non-magnetic material. The interactions between extended dust aggregates are also examined, specifically looking at how the arrangement of charge over the aggregate surface or the inclusion of magnetic material produces dipole-dipole interactions. It will be shown that these dipole-dipole interactions can affect the orientation and structural formation of aggregates as they collide and stick. Analysis of the resulting dust populations will also demonstrate the impact that grain composition and/or charge can have on the structure of the aggregate as characterized by the resulting fractal dimension.

## I. Introduction

UNDERSTANDING the manner in which dust grains coagulate is essential to research across a variety of fields. Astronomical observations have shown the coagulation of dust to be an initial stage of particle growth in the early solar nebula [1] and that grains often grow in size through the formation of fluffy fractal aggregates [2, 3]. Understanding the physics underlying the formation of these aggregates is a key component of being able to accurately model planetary formation, atmospheric processes and coagulation in other environments [4, 5].

The ambient dust environment can have a significant effect on grain coagulation: grains immersed in plasma can become charged, which can either retard or enhance coagulation rates depending on charging conditions [6, 7]. If dust grains in a population charge to a potential of the same sign, the coagulation rate can be reduced or even halt altogether due to the resulting electrostatic repulsion depending on the relative speeds between the grains. In like manner, grain composition can also influence the coagulation rate: both numerical simulations and experimental data have shown that nanometer sized iron grains often coagulate rapidly into small aggregates due to the alignment of their magnetic dipoles [8, 9, 10]. This alignment of the magnetic dipoles creates a weak, attractive force between the grains increasing the probability of contact as one grain approaches another.

This study examines the effects that charged-magnetic grains can have on coagulation, comparing four specific cases: magnetized grains, charged grains, charged-magnetized grains and neutral grains. The results obtained from a numerical simulation will be compared to previous studies on the growth of both charged aggregates [6] and aggregates formed from magnetic materials [8]. The primary purpose of this study is to further examine the aggregation of charged-magnetic grains in order to gain a better theoretical understanding into the basic physics of the aggregation process while also examining the manner in which the grain's charge and magnetization affect this process. These effects will be characterized through the calculation of collision probability, fractal dimension and the morphology of the resultant aggregates.

## II. Method

In this study, aggregates were created by numerically modeling the interactions between colliding particle pairs [11]. Working in the center of mass (COM) frame of an initial seed particle, a second monomer or aggregate was chosen to approach this particle from a random direction. The incoming particle was assumed to have a speed which was a multiple of the thermal velocity $v_{th}$

$$v_{th} = \sqrt{\frac{k_b T}{m}} \quad (1)$$

where $k_b$ is the Boltzmann constant, $T$ is the temperature and $m$ is the particle mass. Charged grains were given velocities of a few times the thermal velocity in order to overcome the Coulomb repulsion, while magnetic grains were given velocities that were only fractions of the thermal velocity.

The forces acting on particle $i$ from the electric or magnetic field of particle $j$ are calculated by

$$\mathbf{F}_{e,ij} = q_i(\mathbf{E}_j + \mathbf{v}_i \times \mathbf{B}_j) \quad (2)$$

$$\mathbf{F}_{m,ij} = (\mathbf{\mu}_j \cdot \nabla)\mathbf{B}_j \quad (3)$$

where $E$ is the electric field, $B$ is the magnetic field, $\mu$ is the magnetic moment, $q$ is the charge and $v$ is the velocity. The charge on an aggregate is approximated using both the monopole and dipole moments [11], while the magnetic dipole moment is calculated as the vector sum of the dipole moments of the individual monomers. The electrostatic and magnetic

dipole moments can also create a torque on each particle, as expressed by:

$$\boldsymbol{\tau}_{ij} = \mathbf{p}_i \times \mathbf{E}_j \quad (4)$$

$$\mathbf{M}_{ij} = \boldsymbol{\mu}_i \times \mathbf{B}_j \quad (5)$$

where $\boldsymbol{\tau}$ denotes the torque due to the electric dipole, $\mathbf{p}$ is the electric dipole moment and $\mathbf{M}$ is the torque due to the magnetic dipole. The resulting torque on each particle can create a rotation of the aggregate, altering its orientation during collision. This in turn can have an effect on the overall fractal dimension of the resulting aggregate.

Collisions are detected when monomers within each aggregate physically overlap. Colliding aggregates are assumed to stick at the point of contact with the orientation of each monomer within the aggregate preserved. Fragmentation during collisions is not considered due to the low velocities imposed on incoming grains [12].

The grains are assumed to be immersed in a plasma environment with the equilibrium charge determined by the balance of the electron and ion fluxes. The charge and electric dipole moment of an aggregate are calculated using a line-of-sight approximation in which electrons and ions are assumed to approach in straight lines from infinity and stick at the point of contact; thus points on an aggregate which have lines of sight blocked by other monomers within the aggregate will have no charging current incident from those directions [7]. This method was used to charge a large number of fractal aggregates. The charge and magnitude of the dipole moment were then plotted vs. $N$, the number of monomers in an aggregate. Results show that the total charge on an aggregate can be well-approximated by a linear log-log fit of the form shown below; the magnitude of the dipole moment can also be estimated by this method, though with less accuracy [7].

$$Q = A\, Q_0\, N^B \quad (6)$$

$$p = C\, p_0\, N^D \quad (7)$$

Here $Q_0$ and $p_0$ act as normalization factors. For monodisperse particles, as used in this study, $Q_0 = Nq_0$, with $q_0$ the (fixed) charge on a given monomer and $p_0$ is the magnitude of the electric dipole moment calculated for a total charge $Q_0$ with the charge arranged such that monomers furthest from the COM have the greatest fraction of the charge [6]. It should be noted that the magnitude of the dipole moment is not well-correlated with $N$, nor is it well-correlated with the fractal dimension, $F_d$. Instead it is most strongly dependent on the geometry of the aggregate. This effect is to be addressed in future work.

Aggregate populations were assembled in three stages. First generation aggregates were grown until $N = 20$ using single monomers. Second generation aggregates were grown until $N = 200$ by modeling collisions between first generation aggregates. Third generation aggregates were grown to a size consisting of approximately $N = 2500$ monomers by modeling collisions between first and second-generation aggregates.

III. INITIAL CONDITIONS

As mentioned above, four initial populations of monomers were examined: 1) magnetic grains, 2) charged grains, 3) charged-magnetic grains, and 4) uncharged grains with no magnetic moment. In each simulation the size and density of the constituent monomers were assumed to be equivalent to that of iron grains of radius $r = 20$ nm and mass $m = 2.66 \times 10^{-19}$ kg. Both magnetic and charged-magnetic grains were given a magnetization of $\mu = 7 \times 10^{-18}$ A·m².

The velocities of incoming particles were set according to the thermal velocity, where $T = 100$ K. Incoming velocities for magnetic grains ranged from the thermal velocity to $v_{th}/16$ in order to match the initial conditions from a previous model [9]. For charged and charged-magnetic simulations, speeds ranged from $2v_{th} \leq v \leq 5v_{th}$. These higher speeds were required to overcome the Coulomb repulsion barrier. An incoming grain's approach vector was directed toward the origin plus an offset or impact parameter, $b$, which varied in magnitude from zero to $3r_{max}$, where $r_{max}$ is the radius of the sphere centered at COM which just encloses the target aggregate.

The plasma temperature was set to give a potential on monomers of $V = -0.0707$ V; the aggregate charging model yielded fits for charge and electric dipole moment given by

$$Q = 0.98\, N^{0.50}\, q_0 \quad (8)$$

$$|p| = 0.10\, N^{-0.12}\, |p_0| \quad (9)$$

IV. RESULTS

A. Collision Probability

Collision statistics for each potential interaction were collected including collision outcome, initial velocity, impact parameter, and the number of monomers ($N$) in the resulting aggregate. These statistics were then used to determine the probability of interaction and coagulation between particles under various initial conditions for each dust population.

The dipoles of magnetized grains tended to align as they approached each other; upon collision the dipole moment was assumed to be "frozen in" contributing to the total magnetic dipole moment of the entire aggregate. Figure 1 shows the normalized magnetic moment of a cluster, $\mu/\mu_0$, where $\mu_0$ is the magnetization of a single grain, as a function of aggregate size. The trend line indicates the magnetic moment increases as $\mu \propto \mu_0 N^{0.53}$, similar to previously published results [8]. This is in contrast to the electric dipole moment, which is only weakly dependent on $N$. The magnetic dipole moment tends to grow with the addition of each monomer or aggregate, as the magnetic dipoles tend to align independent of the point of collision. The electric dipole, on the other hand, is determined by the plasma fluxes to unblocked surfaces of monomers within an aggregate, and thus depends most strongly on the final geometry of the aggregate.

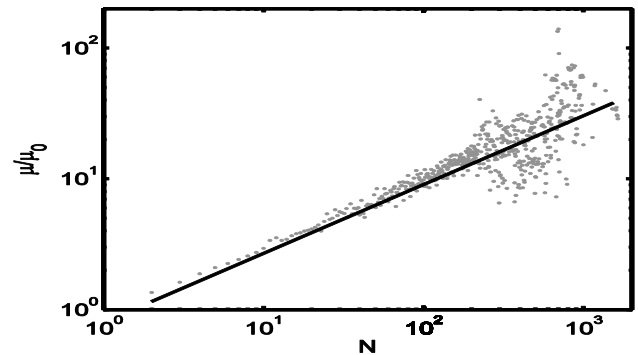

Figure 1. Normalized cluster magnetic moment as a function of $N$, the number of monomers in an aggregate. The fit line shows an exponential increase $\mu \propto N^{0.53}$.

Collision probability statistics were taken for relative velocities between particles as shown in Figure 2. Incoming magnetic aggregates (Fig. 2(a)) show a high probability of colliding with the target aggregate for all velocities up to the thermal velocity. At lower speeds, magnetic grains allow alignment of the magnetic dipoles which in turn enhances the interparticle attractive force and therefore the collision probability. As expected, missed collisions occurred most often at higher velocities, where the probability of dipole alignment is at a minimum.

As mentioned above, both charged and charged-magnetic grains were given higher velocities in order to overcome the Coulomb repulsion. For the size regimes of grains modeled, the minimum collisional velocity was determined to be approximately $3v_{th}$. Below this threshold, both grain types showed nearly the same collision probability. From Fig. 2(b), it can be seen that charged-magnetic grains exhibit a higher probability of coagulation at speeds greater than this, evidently due to coagulation being enhanced by the addition of the additional attractive forces created by the magnetization. The maximum difference between collision probabilities for the two populations is approximately 12% at speeds between $3v_{th}$ and $4v_{th}$. At the highest speeds examined (above $4.5v_{th}$) the collision probabilities become similar once again, approaching a limiting value of one. At those high speeds the collisions between grains are similar to ballistic collisions.

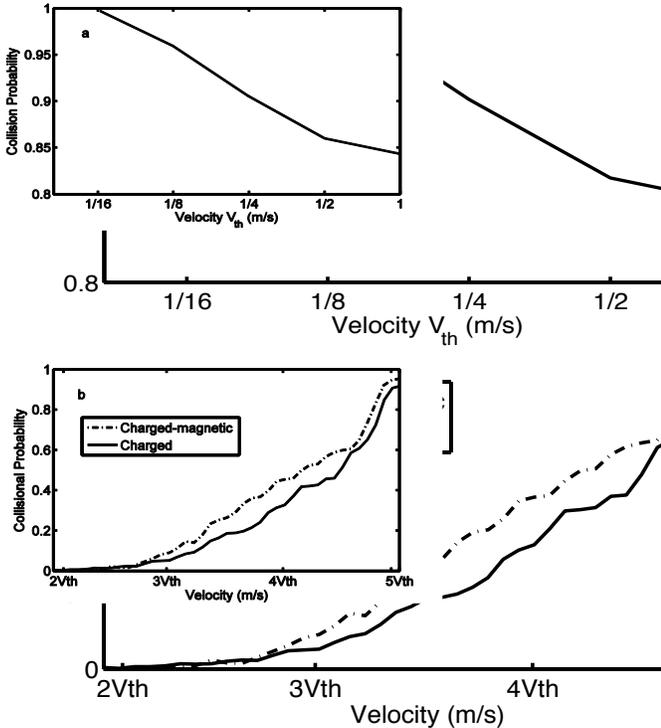

Figure 2. a) Probability for collision of magnetic grains having an initial speed of $1/16v_{th} \leq v \leq v_{th}$. b) Probability of collision for charged and charged-magnetic aggregates having an initial speed of $2v_{th} \leq v \leq 5v_{th}$.

When examining collision probabilities as a function of impact parameter, the difference between charged and charged-magnetic grains is less pronounced (Fig. 3). Impact parameters with values less than $r_{max}$, yield an appreciably higher collision probability for charged-magnetic grains as compared to charged grains. However, for larger impact parameters both grain types exhibit similar collision probabilities, with an only slightly enhanced collision probability for charged-magnetic grains. This is in large part due to the short-range nature of the dipole-dipole interactions. Neutral grains undergoing ballistic collisions were examined and seen to have a collision probability near one for impact parameters up to the grazing distance, the point at which the radii of the grains barely overlap, and zero beyond that. This is in large part due to the compact nature of the aggregates formed from this population, as discussed in the next section.

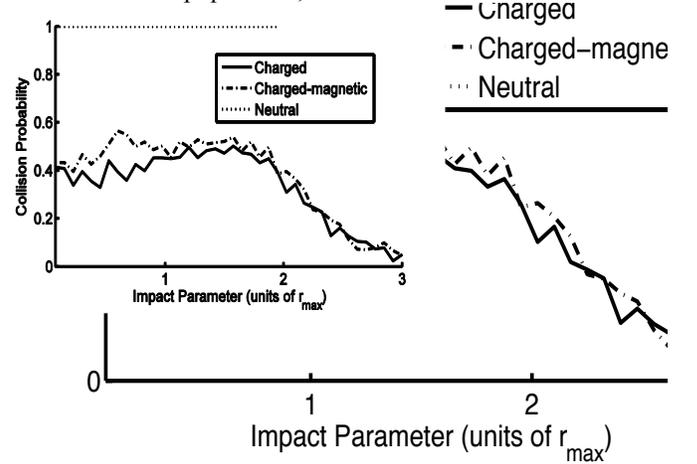

Figure 3. Collision probability for charged, charged-magnetic, and neutral grains vs. impact parameter.

### B. Fractal Dimension

The fractal dimension provides a measure of the openness or porosity of a structure. It can range from nearly one for a linear structure, to three for a compact sphere. The fractal dimension is thus an important parameter for collisional processes since fluffy, extended aggregates will obviously exhibit a much higher collisional cross-section. At the same time, open porous structures are in general entrained in the gas in their environment, which can suppress the relative velocities between grains, reducing coagulation rates.

In this study, the fractal dimension is found by enclosing an aggregate within a cube of side $a$ divided into $a_0^3$ subboxes. The number of subboxes containing a portion of the aggregate is given by $N(a_0)$ allowing the fractal dimension to be calculated as

$$F_d = \frac{\log N(a_0)}{\log(a_0)} \qquad (10)$$

Aggregates assembled from a magnetic material tend to form filamentary structures consisting of many linear chains. Such structures have been created and observed in laboratory experiments [13]. Magnetic grains tend to form these chains through the partial or total alignment of the dipole moments of local chains. Similar linear structures can be seen within the typical small aggregate formed from these numerical simulations as shown in Fig. 4(a). Larger aggregates composed of magnetic grains are less linear; however the total structure still consists of linear chains as most easily seen at the periphery of the aggregate (see Fig. 4(b)). A greater induced spin on an incoming monomer or aggregate reduces the dipole alignment in the resultant aggregate. For magnetic populations this also reduces the linear nature of any resulting aggregate structures.

Charged monomers tend to form a denser, more compact structure as can be seen in Fig. 4(c, d). Charged-magnetic grains exhibit a behavior intermediate to both magnetic and charged aggregates, as expected (Fig. 4(e, f)). While branches are more clumped than those for magnetic aggregates, they are decidedly more linear than an equivalent non-magnetic charged structure (Fig. 4(d)). Neutral grains form relatively dense and more spherical aggregate structures at all sizes, as shown in Fig. 4(g, h).

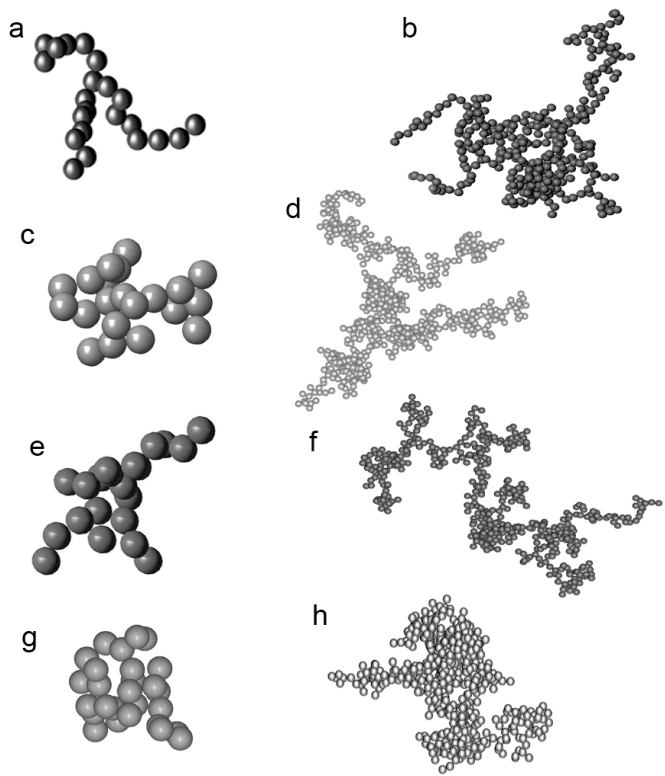

Figure 4. Comparison of aggregates formed from the following populations: Magnetic aggregates a) $N = 22$, $F_d = 2.159$, b) $N = 334$, $F_d = 1.789$; Charged aggregates c) $N = 21$, $F_d = 2.333$, d) $N = 595$, $F_d = 1.907$; Charged-magnetic aggregates e) $N = 21$, $F_d = 2.359$, f) $N = 504$, $F_d = 1.819$; and Neutral aggregates g) $N = 25$, $F_d = 2.403$, h) $N = 589$, $F_d = 1.976$.

The fractal dimension for each population of aggregates as a function of $N$ is shown in Fig. 5. The best-fit lines are shown only including data for aggregates with $N \geq 10$. The smallest aggregates ($N < 10$) are omitted from the calculation to produce these fit lines due to a lack of variation in the fractal dimension between populations at small $N$. A general decrease in fractal dimension can be seen as $N$ increases, as large aggregates have a more porous structure than do small aggregates, which consist of only a few spherical monomers. Uncharged aggregates formed from magnetic material exhibit the lowest fractal dimension, as expected. It is interesting to note that while the charged grains and charged-magnetic grains have nearly identical fractal dimensions for the smaller aggregates ($N \approx 10$), the charged-magnetic aggregates tend towards lower fractal dimensions for larger $N$, exhibiting an increasingly open structure. This open structure is due to the existence of local linear chains of monomers in the magnetic aggregate structures. These linear chains are formed through the alignment of the magnetic dipole moments during coagulation for a single incoming monomer. This indicates that the magnetic force between aggregate structures becomes increasingly important relative to the electrostatic interactions, largely due to the fact that the magnetic moment of the aggregates increases with $N$ (c.f. Figure 1) while the electrostatic dipole moment is only weakly dependent on $N$. Aggregates constructed from ballistic collisions (assuming no interaction forces between grains) tended to have the largest fractal dimensions at all sizes.

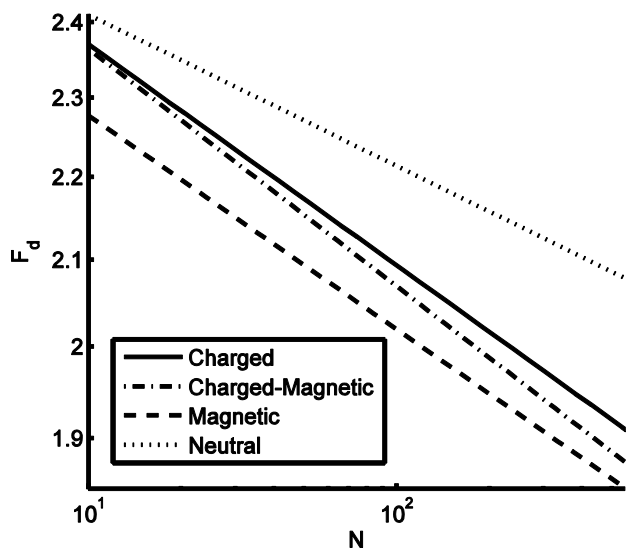

Figure 5. A comparison of the average fractal dimension for each population of aggregates as a function of size. Each line is a best fit to the fractal dimension of all of the aggregates with $N \geq 10$ in a given population. Error bars have been omitted for clarity.

Probability density estimates for fractal dimension were also calculated for each population as shown in Figure 6. Analysis was done using an equal number of second- and third-generation grains from each population; first generation aggregates were not included because of the similarity of fractal dimensions at low $N$ across all four populations. As shown in Fig. 6 (a), magnetic aggregates have a local maximum at a lower fractal dimension when compared to other populations, centered approximately at $F_d = 2.05$, with a significant percentage, 36.5%, of the aggregates with fractal dimensions less than 2.0. Charged and charged-magnetic grains (Fig. 6 (b,c)) show a much broader range of fractal dimensions. Charged-magnetic grains do, however, have a greater percentage of aggregates with fractal dimensions at the lowest fractal dimensions ($F_d < 2.0$), 36.5% for charged-magnetic and 32.4% for charged aggregates, and a smaller percentage of aggregates with large fractal dimensions ($F_d > 2.2$), 27.1% for charged-magnetic and 30.8% for charged grains, respectively. This indicates the influence of the magnetic interactions on the coagulation process. Neutral grains (Fig. 6 (d)) show a narrow peak near $F_d$ = 2.25, a larger fractal dimension than for any of the other populations. The neutral population has essentially zero aggregates form with the lowest fractal dimensions ($F_d \leq 1.8$).

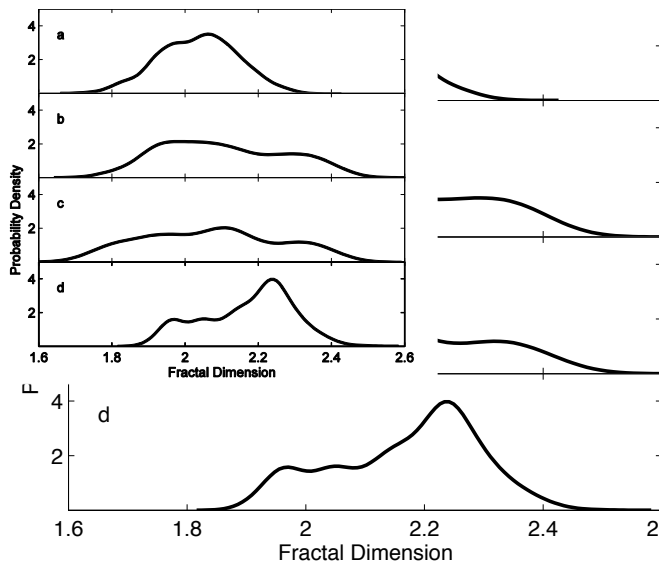

Figure 6. Probability density estimates for the fractal dimensions of each population. a) Magnetic grains exhibit a local maximum at $F_d$ = 2.05. b) Charged aggregates have a broad distribution in fractal dimension. c) Charged-magnetic aggregates also show a broad distribution, but have a greater percentage of aggregates with $F_d$ < 2.0, compared to the charged grains. d) Neutral aggregates have a relatively narrow peak at $F_d$ = 2.25.

V. CONCLUSIONS

A comparison between aggregate structures resulting from collisions between magnetic and non-magnetic, charged and uncharged dust populations has been given. Initial simulations suggest that aggregate populations assembled from charged-magnetic grains exhibit behavior intermediate to that shown by aggregates assembled from purely charged or magnetic grains, as expected. This behavior can be clearly seen in both the collision probabilities between aggregates as shown in Figures 2 and 3, and the fractal dimension of the resulting aggregates as seen in Figures 4 - 6.

The collision probability for populations grown from charged grains and charged-magnetic grains leads to the conclusion that the charge on the grains dictates behavior of the aggregates, especially for small $N$. On the other hand, magnetization of the grains provides an attractive force between aggregates which can lead to a higher probability of collision. This difference is most pronounced for low velocities (given velocities large enough to overcome the Coulomb repulsion barrier) and small impact parameters. Magnetic forces play an increasingly important role in determining the fractal dimension for large $N$ as the magnitude of the magnetic moment grows larger with $N$ (Fig. 1) while the electrostatic dipole moment is only weakly dependent on $N$.

The results of this study imply that the characteristics of aggregate growth for a given environment are controlled by both the material properties (for instance ferrous materials will experience magnetic interactions) as well as the plasma parameters which determine the charge on individual monomers. This not only has significant implications for the early stages of planetesimal formation, which will be dependent upon the local conditions within a protoplanetary disk, but also for coagulation in a laboratory environment, where the structure and size of the aggregates might be able to be controlled by adjusting the material properties and plasma characteristics.

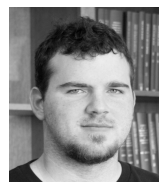

**Jonathan D. Perry** was born in Dallas, TX in 1986. He received the B.A. degree in physics from Baylor University in Waco, TX in 2009.
He is currently pursuing a Ph.D. in physics at Baylor University.

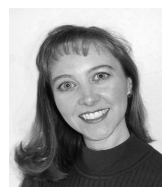

**Lorin S. Matthews** was born in Paris, TX in 1972. She received the B.S. and the Ph.D. degrees in physics from Baylor University in Waco, TX, in 1994 and 1998, respectively.
She is currently an Assistant Professor in the Physics Department at Baylor University. Previously, she worked at Raytheon Aircraft Integration Systems where she was the Lead Vibroacoustics Engineer on NASA's SOFIA (Stratospheric Observatory for Infrared Astronomy) project.

**Truell W. Hyde** was born in Lubbock, Texas in 1956. He received the B.S. in physics and mathematics from Southern Nazarene University in l978 and the Ph.D. in theoretical physics from Baylor University in 1988.

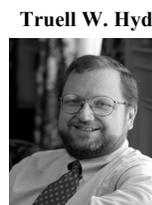

He is currently at Baylor University where he is the Director of the Center for Astrophysics, Space Physics & Engineering Research (CASPER), a Professor of physics and the Vice Provost for Research for the University. His research interests include space physics, shock physics and waves and nonlinear phenomena in complex (dusty) plasmas.